\documentclass[pdflatex,sn-mathphys-num]{sn-jnl}

\usepackage{graphicx}
\usepackage{amsmath,amssymb,amsfonts}
\usepackage{cancel}
\usepackage{float}
\usepackage{xcolor}
\usepackage{epstopdf}

\newcommand{\ac}[1]{\textcolor{black}{#1}}

\newcommand{\pdern}[3]{\frac{\partial^#3#1}{\partial#2^#3}}
\newcommand{\pder}[2]{\frac{\partial#1}{\partial#2}}

\title{\ac{Breathers across the Focusing-to-Defocusing Transition: From Diffraction-Like Dynamics to Talbot Carpets}}

\author*[1]{\fnm{Andrea} \sur{Armaroli}}\email{andrea.armaroli@unife.it}
\equalcont{These authors contributed equally to this work.}
\author*[2]{\fnm{Hung-Chu} \sur{Hsu}}\email{hchsu@mail.nsysu.edu.tw}
\equalcont{These authors contributed equally to this work.}
\author[3]{\fnm{Meng-Syue} \sur{Li}}
\author[4]{\fnm{Lili} \sur{Kimmoun}}
\author[5]{\fnm{Jie} \sur{Zhang}}
\author[6,7]{\fnm{Michel} \sur{Benoit}}
\author[8]{\fnm{Nobuhito} \sur{Mori}}
\author[1]{\fnm{Stefano} \sur{Trillo}}
\author*[9,10,11]{\fnm{Amin} \sur{Chabchoub}}\email{amin.chabchoub@oist.jp}

\affil[1]{\orgdiv{Department of Engineering}, \orgname{University of Ferrara}, \postcode{44122}, \city{Ferrara}, \country{Italy}}
\affil[2]{\orgdiv{Department of Marine Environment and Engineering}, \orgname{National Sun Yat-Sen University}, \city{Kaohsiung}, \postcode{80424}, \country{Taiwan}}
\affil[3]{\orgname{National Academy of Marine Research}, \city{Kaohsiung}, \postcode{806614}, \country{Taiwan}}
\affil[4]{\orgname{Aix-Marseille Univ, CNRS, Centrale M\'editerran\'ee, IRPHE}, \postcode{13013}, \city{Marseille}, \country{France}}
\affil[5]{\orgdiv{College of Shipbuilding Engineering}, \orgname{Harbin Engineering University}, \city{Harbin}, \postcode{150001}, \country{China}}
\affil[6]{\orgname{EDF R\&D, Laboratoire National d'Hydraulique et Environnement (LNHE)}, \postcode{78400}, \city{Chatou}, \country{France}}
\affil[7]{\orgname{LHSV -- Saint-Venant Hydraulics Laboratory, ENPC, Institut Polytechnique de Paris, EDF R\&D}, \postcode{78400}, \city{Chatou}, \country{France}}
\affil[8]{\orgname{Disaster Prevention Research Institute, Kyoto University}, \city{Kyoto}, \postcode{611-0011}, \country{Japan}}
\affil[9]{\orgname{Marine Physics and Engineering Unit, Okinawa Institute of Science and Technology}, \city{Okinawa}, \postcode{904-0495}, \country{Japan}}
\affil[10]{\orgdiv{Department of Civil and Environmental Engineering}, \orgname{Imperial College London}, \postcode{SW7 2AZ}, \city{London}, \country{United Kingdom}}
\affil[11]{\orgdiv{Department of Infrastructure Engineering}, \orgname{The University of Melbourne}, \state{Victoria}, \postcode{3010}, \country{Australia}}

\begin{document}
\maketitle
\begin{abstract}

We experimentally investigate fundamental unstable breather wave groups propagating across the transition from focusing to defocusing regimes. In hydrodynamics, this transition is realized through a rapid reduction in water depth over a short spatial interval. We demonstrate that the time-periodic Akhmediev breather develops a Talbot-like recurrence pattern in the defocusing regime, although only its initial stages are experimentally accessible. In contrast, the Peregrine breather exhibits qualitatively different behavior, characterized by oscillatory envelope dynamics. These oscillations can be understood as a purely linear phenomenon arising within the nonlinear Schrödinger framework when dispersion dominates over nonlinearity, as further confirmed by numerical simulations over extended propagation distances. The experimental observations agree closely with finite-depth nonlinear Schrödinger predictions across regimes with substantially different relative strengths of dispersion and nonlinearity. Our findings reveal a temporal analogue of diffraction and provide a foundation for exploring focusing-to-defocusing transitions in other nonlinear dispersive media.
\end{abstract}

\section*{Introduction}
Transitions between different propagation regimes are ubiquitous in physical systems. Nonlinear hydrodynamic waves are particularly sensitive to variations in water depth, which play a critical role in coastal hydrodynamics by governing processes such as wave shoaling and run-up \cite{Carrier_Greenspan_1958,Li2002,Sriram2014,li2023formation}; nonlinear wave interactions radically modify wave statistics \cite{trulsen2012laboratory,zhang2019statistics,li2021rogue,lawrence2021statistical,karmpadakis2022new,mendes2026energetics}, while variations in group-velocity dispersion and the Kerr nonlinear coefficient strongly influence the dynamics of solitary waves in nonlinear optics \cite{anastassiou1998optically,liang2019transition}. Thus far, only a few theoretical and experimental studies have investigated the evolution of modulated wave packets such as solitons and breathers in non-uniform media \cite{Chernikov1991,Chernikov1992,Abdullaev2005,Armaroli2020,gomel2021stabilization}.

A key step towards understanding the effects of regime transitions on nonlinear waves can be obtained by studying breathers, beating wave groups describing the nonlinear stage of modulational instability (MI) of waves propagating in a uniform medium and strong candidates for modeling rogue waves \cite{onorato2013rogue,dudley2014instabilities,tlidi2022rogue}. Within the nonlinear Schr\"odinger equation (NLSE) framework \cite{zakharov1968stability,shabat1972exact}, two fundamental solutions occurring in the focusing regime, i.e., the Akhmediev breathers (AB) \cite{akhmediev1985generation} and the doubly-localized Peregrine breather (PB) \cite{peregrine1983water}, have been extensively studied with a striking agreement between theory and observations \cite{kibler2010peregrine,chabchoub2011rogue,bailung2011observation,romero2024experimental} and have been shown to be the fundamental building blocks to trigger and control complex dynamical phenomena in a variety of nonlinear dispersive systems \cite{mussot2018fibre,Pierangeli2018,luo2020creation,suret2020nonlinear,suret2024soliton}.

Here, we report an experimental study analyzing the propagation of modulationally unstable wave groups in the form of ABs and PBs over a rapid depth variation in a large-scale water wave facility, which implies a transition from the focusing to the defocusing regime, i.e., a sign change of the nonlinearity in the hydrodynamic NLSE \cite{mei2005theory,chabchoub2015nonlinear}. Our data allow us to explore the long-term behavior of these two types of fundamental wave groups before and after the transition. \ac{The experiments show that ABs exhibit the initial stages of Talbot-like pattern formation upon transitioning into a nearly linear regime}, while PBs transform into an oscillating pattern that spreads while propagating. This can be interpreted as a time-analogue of diffraction, since linear effects dominate over nonlinear ones. This latter phenomenon is sensitive to the phase of the wave packet at the point where nonlinearity ceases to be effective. These findings are expected to stimulate further experimental, numerical, and theoretical work, not only in water waves, but also in optics, plasma, transmission lines, and Bose-Einstein condensates, to elucidate which linear and nonlinear effects are triggered by rapid dispersion and nonlinearity variations in the medium.

\section*{Results}
\begin{figure}[h]
\centering
\includegraphics[width=.95\linewidth]{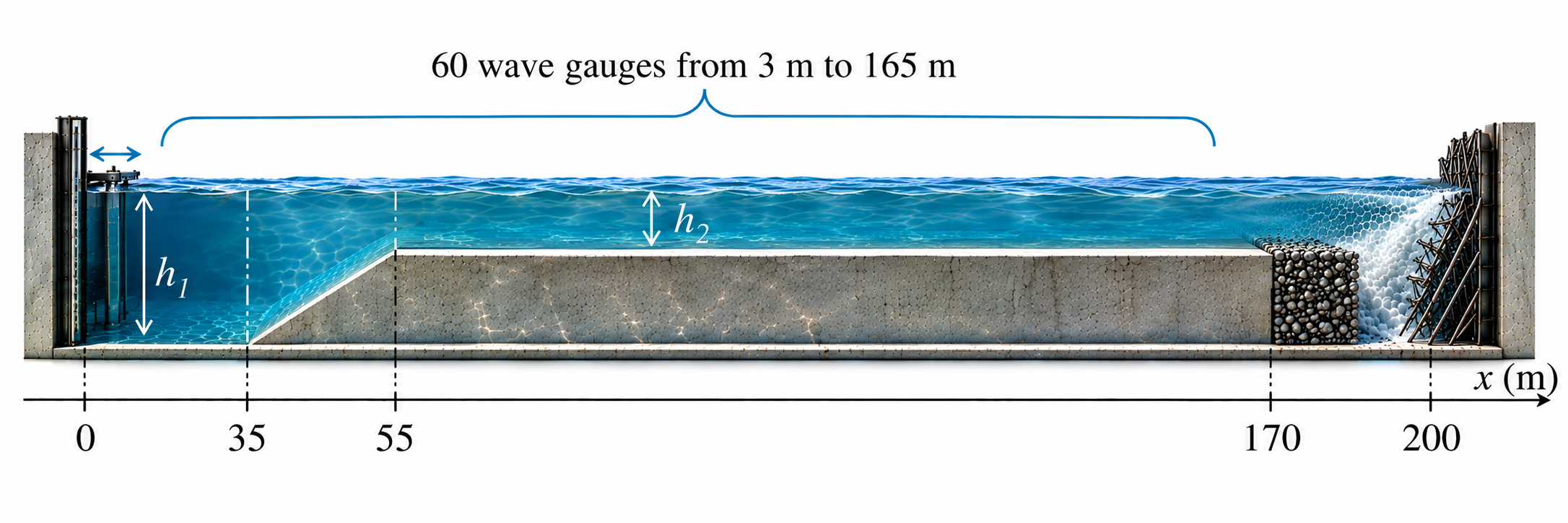}
\caption{Large-scale unidirectional water wave tank schematic, depicting the piston-type wave maker, the wave gauge locations, the bathymetry characteristic change, and the wave absorption installations. The initial depth is $h_1=1.2$~m for AB initial conditions (respectively, $h_1=1.3$~m for the PB) and becomes $h_2=0.2$~m (respectively, $h_2=0.3$~m) after the 1/20 slope transition.}\label{fig1}
\end{figure}

We have conducted numerous experiments on AB and PB transitions across a range of breather and carrier wave parameters, but here we present only four representative cases, two for each family of breather boundary conditions, to illustrate the underlying physical phenomena investigated in this study. 
The experiments have been performed in the large-scale facility of the Tainan Hydraulics Laboratory (THL), Taiwan, with dimensions 200$\times$2$\times$2 m$^3$, suitable for studying coastal processes and tracking the long-term evolution of hydrodynamic waves. The long-crested waves are generated by a hydraulic piston-type wave maker, suitable for generating waves in finite water depth. An artificial concrete bathymetry was installed to emulate a rapid depth transition from $h_1=1.2$~m to $h_2=0.2$~m for the AB wave packet (respectively, $h_1=1.3$~m to $h_2=0.3$~m for the PB wave packet) through a linear transition over a distance of 20~m, resulting in a $1/20$ slope. Moreover, 60 resistive wave gauges have been deployed along the facility to track the wave packet progression in the different depth zones. Details of the experimental setup are summarized and shown in Fig.~\ref{fig1}. This bathymetric transition can be considered rapid, not relative to the carrier wavelength (of the order of one meter), but rather to the characteristic length scale of nonlinear evolution described by the FDNLSE, which exceeds 20~m for the typical initial carrier wave parameters considered here.  

Assuming a rapid, linear bathymetric transition from an initial depth $h_1$ to a final depth $h_2$, the wavenumber $k$ along the wave tank can be determined from the linear surface gravity wave dispersion relation, $\omega^2 = gk\tanh(kh)$, where $\omega = 2\pi f$ is the angular frequency and $g$ denotes gravitational acceleration. In order to model the evolution of the envelope along the tank, we employ the finite-depth NLSE (FDNLSE), where the coefficients depend explicitly on the product $\kappa=k h$ and shoaling is neglected (see \hyperref[sec:methods]{Methods}). Importantly, at $\kappa=1.363$, the nonlinearity vanishes, allowing for a transition from the focusing to defocusing regime \cite{carretero2024nonlinear}. The boundary conditions, i.e., the respective wave packets as generated by the wave maker, are determined by the parametrization of exact FDNLSE solutions and the carrier wave parameters, e.g., wave frequency and wave steepness, so that the maximal theoretical focusing is expected to occur near the junction between focusing and defocusing regimes, that is, when $\kappa=1.363$ is satisfied on the slope. 
Note that the wave envelope is extracted from the measured surface elevation data using the Hilbert transform, enabling a direct comparison between the observed wave evolution and theoretical predictions based on the FDNLSE \cite{osborne2010nonlinear,kimmoun2016modulation}.

\begin{figure}[h]
\centering
\includegraphics[width=1.0\linewidth]{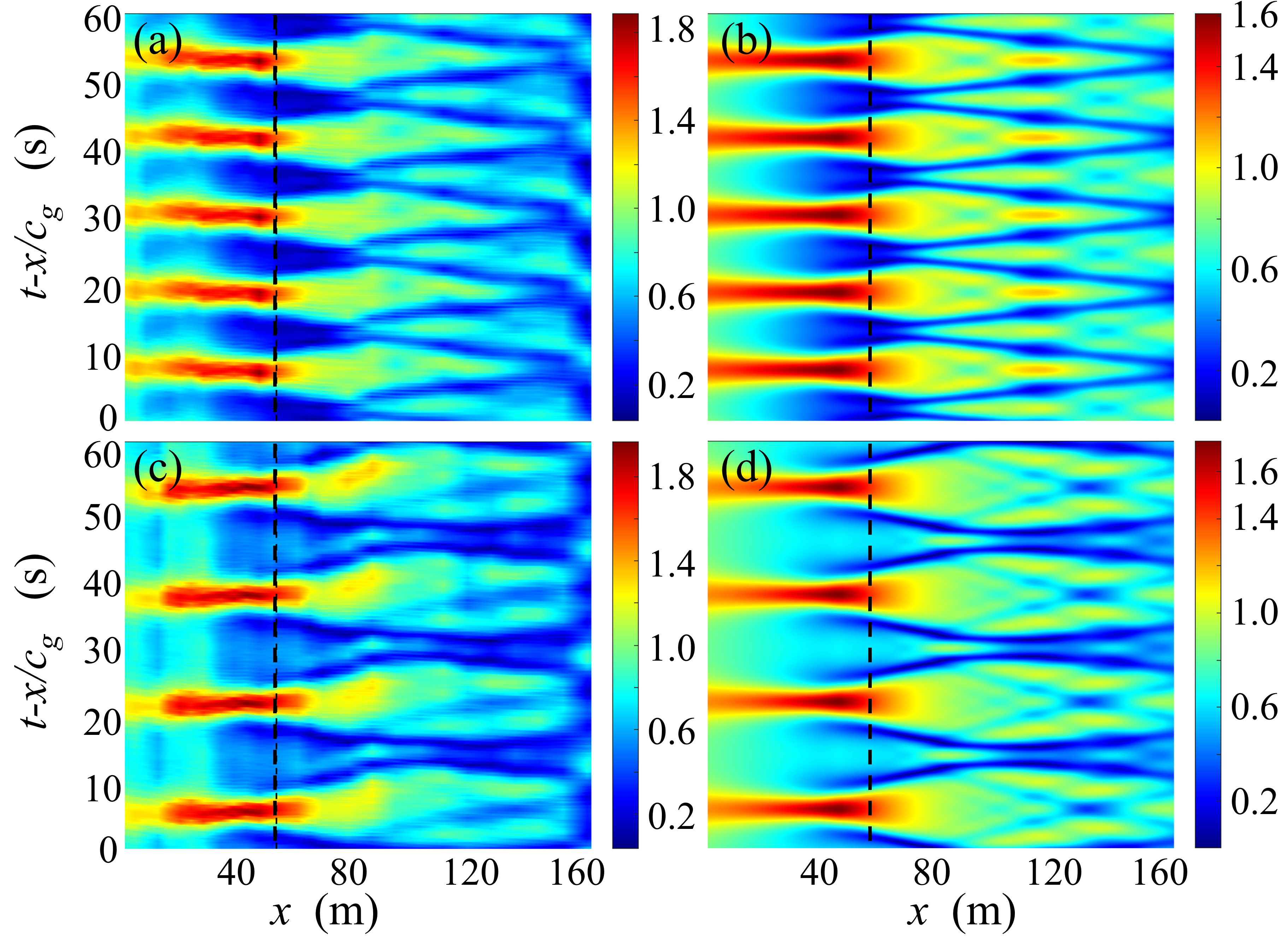} 
\caption{Evolution of two AB envelopes normalized by the initial carrier amplitude as propagating over the variable bathymetry in the wave flume. Left: Experiments. Right: Corresponding FDNLSE simulations. (a) and (b) AB modeling the case of maximal growth rate for $\mathfrak{a}=0.25$ for the carrier steepness $\varepsilon=0.1$ and frequency $f=1$ Hz.
(c) and (d) AB modeling the case of higher-order MI for $\mathfrak{a}=0.37$ and adopting the same carrier parameters as in the previous case. The vertical dashed line denotes the location of the transition at $\kappa=1.363$ or $\gamma=0$.
}
\label{fig2}
\end{figure}

Fig.~\ref{fig2} illustrates the evolution of the two AB wave envelopes propagating in the wave flume with the rapid change in water depth over the variable bathymetry. The first describes the case of classical modulation instability dynamics at its maximal growth rate, i.e., following Ref.~\cite{kimmoun2016modulation}, $\mathfrak{a}=0.25$, while the second is for a larger perturbation period, close to the center of the MI frequency range,  which corresponds to $\mathfrak{a}=0.37$, i.e., the limit where higher-order modulation instability appears \cite{erkintalo2011higher,kimmoun2017nonconservative}. 

The carrier wave parameters are provided in the figure caption. Note that the AB envelopes are parameterized as in \cite{kimmoun2016modulation} and are recalled in \hyperref[sec:methods]{Methods}. From these measurements, we draw three conclusions: first, we can clearly observe that unstable waves keep their coherence as well as their focusing trend on the slope; second, the rogue-like wave patterns do not immediately disappear when entering the defocusing regime; third, when propagating further in the tank starting from the initial $\kappa_1=4.80$ into a shallower regime satisfying $\kappa_2=1.02$, the wave packets first broaden, then break into a characteristic periodic pattern. We will discuss below the long-term evolution beyond experimental constraints. Both measurements are in very good agreement with the FDNLSE simulations, accounting for the same bathymetry change; the dissipation coefficient has been determined from the measured wave attenuation and is set to $\mathfrak{D}=3.8\times 10^{-3}$ m$^{-1}$. The phenomenology in the defocusing regime exhibits for both values of $\mathfrak{a}$ a pattern of finer and finer fringes with the successive appearance of secondary refocusing events. This hints at a phenomenology similar to Talbot carpets \cite{Berry1996Talbot} in diffraction theory, where the fringes generated by a diffraction grating interact constructively or destructively, ideally producing a self-image of the grating itself at well-predicted lengths. \ac{However, despite the considerable length of the wave facility, only the initial stages of these dynamics could be observed experimentally. Their subsequent evolution is therefore investigated through numerical simulations extending beyond the experimental domain under identical physical parameters, as discussed below.} 

\begin{figure}[tb]
\centering
\includegraphics[width=1.0\linewidth]{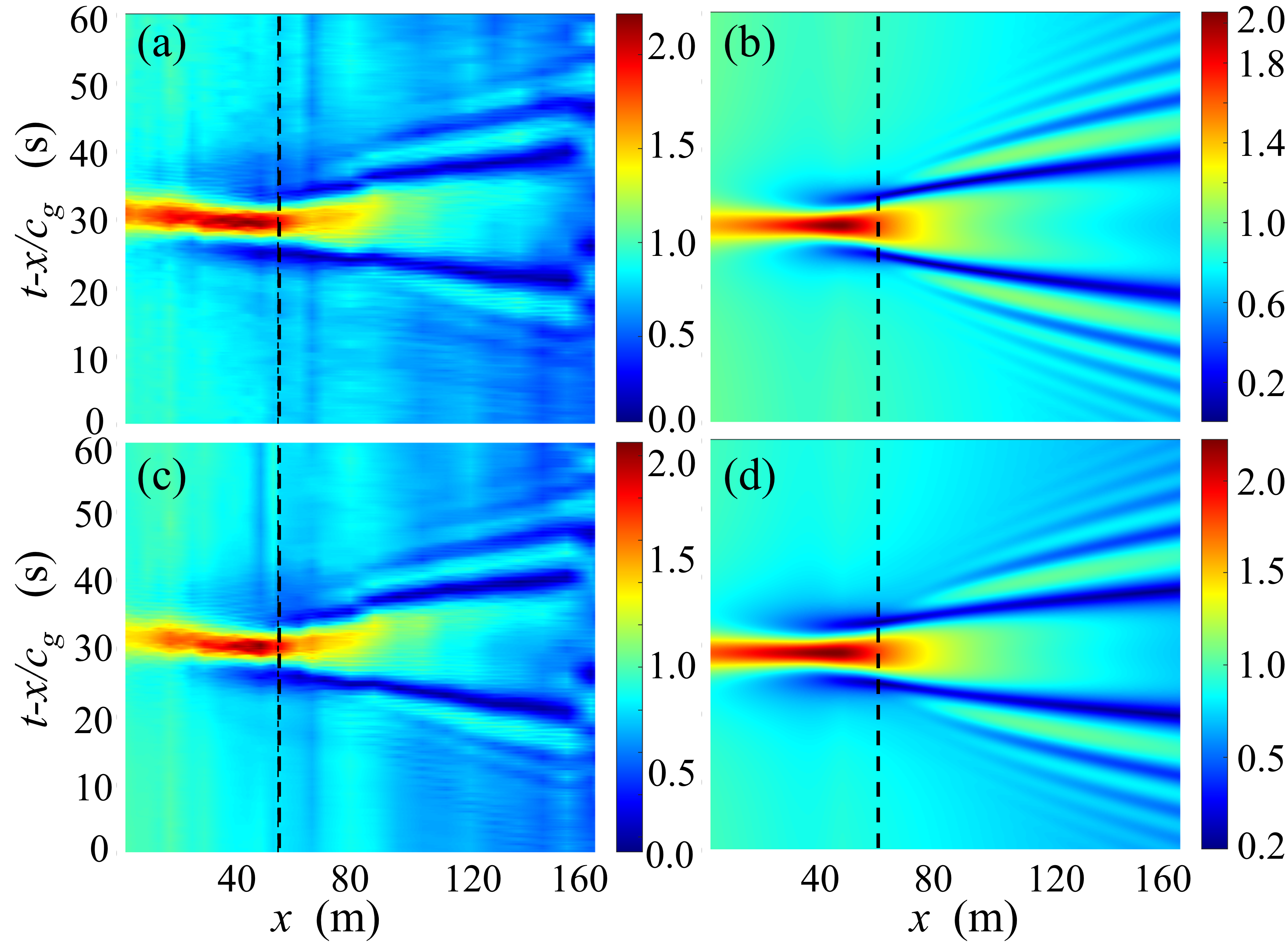}
\caption{Evolution of two PB envelopes normalized by the initial carrier amplitude as propagating over the variable bathymetry in the wave flume. The carrier frequency is $f=0.83$ Hz. Left: Experiments. Right: Corresponding FDNLSE simulations. (a) and (b) steepness $\varepsilon=0.09$. (c) and (d)  steepness $\varepsilon=0.11$. 
The vertical lines share the same meaning as in Fig.~\ref{fig2}.}\label{fig3} 
\end{figure}

Next, we focus on the rapid depth-transition dynamics of the PB, which features  distinct and interesting dynamics. The experimental results of two runs are shown in Fig.~\ref{fig3}. We present two slightly different initial steepness values, as reported in the caption. 
While the initial focusing process in the deeper part of the tank is similar to the time-periodic AB observations, the evolution of the PB differs after transitioning into the shallow area. In these experiments, the depth transition is from $\kappa_1=3.62$ to $\kappa_2=1.05$. We clearly observe that an oscillating and diverging pattern emerges. The PB is characterized by two symmetric dips close to the rogue event: similar fission wave dynamics has been suggested in a numerical work in which the PB is propagating in the presence of shear currents of slowly varying vorticity \cite{hsu2021transformation}. In simulations, Fig.~\ref{fig3}(b,d), where the dissipation coefficient is set to $\mathfrak{D}=3.6\times 10^{-3}$ m$^{-1}$, it is apparent instead that the two dips tend to spread during propagation. Between these two depressions, the PB peak broadens to an almost uniform background, gently decaying during evolution as a result of overall dissipation. It is worth noting how the pattern develops in propagation on the two temporal sides of the envelope: an oscillating pattern alternating peaks and dips is clearly visible. A closer inspection shows that this pattern does not diverge with a regular speed, as could be expected from a dispersive shock wave (DSW) fan modeled by the Korteweg-de Vries-type or Whitham evolution equations \cite{zabusky1965interaction,fatome2014observation,trillo2016experimental,el2016dispersive}. 
\ac{This evolution can therefore be interpreted as a linear diffraction-like phenomenon, as wave damping and the resulting weakening of nonlinear effects prevent the onset of the DSW regime within the experimentally accessible propagation distance.} This will be further justified by an approximate linear theory presented in the \hyperref[sec:methods]{Methods} section. 

\begin{figure}
    \centering
    \includegraphics[width=1.0\linewidth]{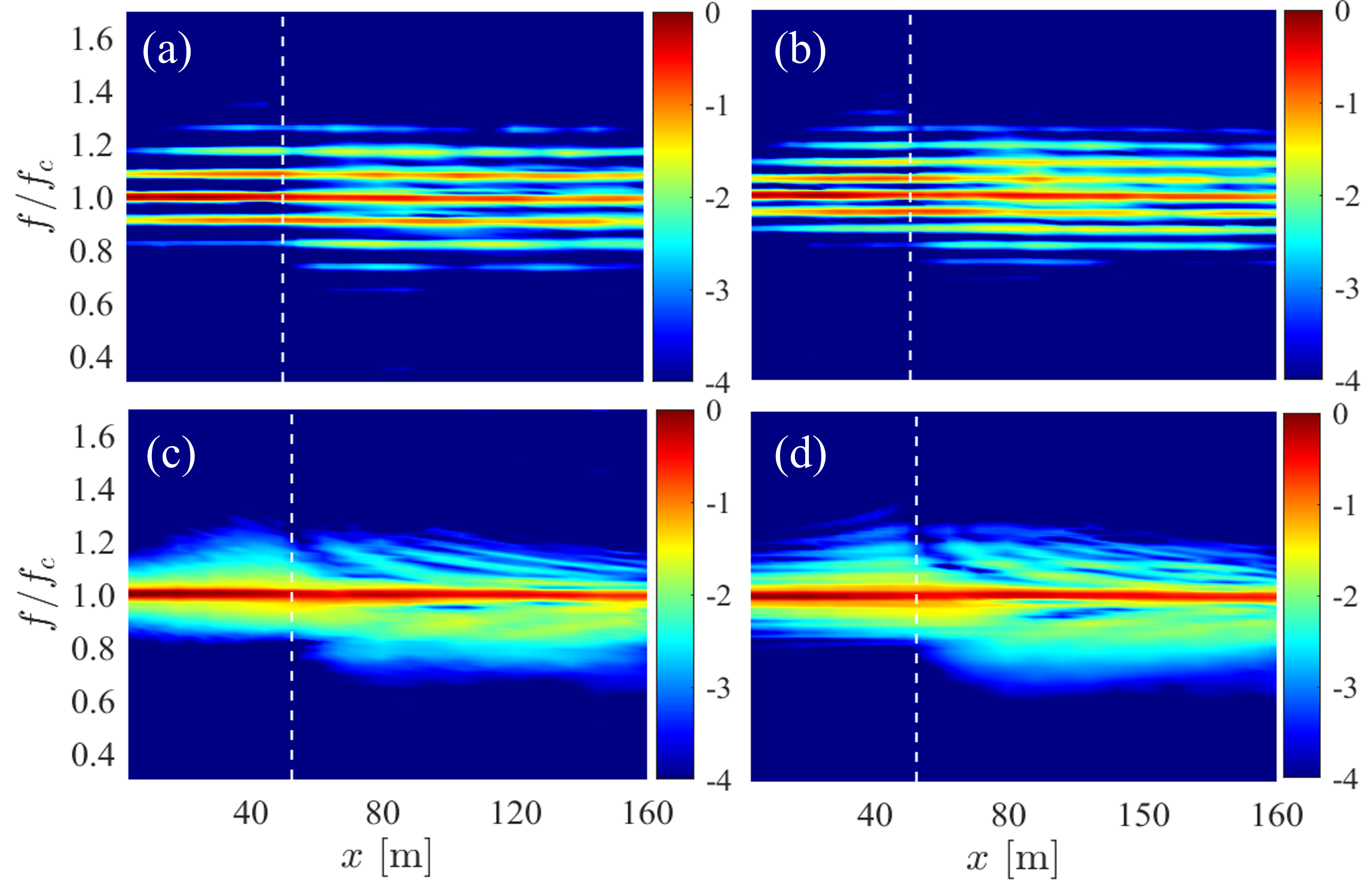}
    \caption{Evolution of the respective normalized frequency spectra $\textrm{log}_{10}[S(f)/\textrm{max}(S(f))]$: (a) the case of Fig.~\ref{fig2}(a), (b) the case of Fig.~\ref{fig2}(c), (c) the case of Fig.~\ref{fig3}(a), (d) the case of Fig.~\ref{fig3}(c). The vertical lines share the same meaning as in Fig.~\ref{fig2}.}
    \label{fig4}
\end{figure}

The corresponding spatial evolution of the normalized spectra of these cases is shown in Fig.~\ref{fig4}.
Figs.~\ref{fig4}(a,b) show the evolution of the AB spectra. As it is well understood, the evolution in the deeper region is characterized by the cascaded generation of phase-locked frequency components: apart from the carrier, two pairs of sidebands are clearly identified. In Fig.~\ref{fig4}(b), they are more closely packed, as expected from the larger value of the breather parameter $\mathfrak{a}$. As soon as the wave propagates in the defocusing regime, this process is almost halted. We observe weak oscillations in space that account for a quasi-periodic conversion between the carrier and the sideband pairs, as occurs outside of  MI sidebands \cite{Vanderhaegen2021}. 
In contrast, the PB exhibits qualitatively different behavior, with an initial nearly symmetric spectral broadening in the focusing regime followed by additional asymmetric broadening. Within the limitations of spectral reconstruction from noisy time traces, this looks more apparent for the larger steepness, so it may be of nonlinear origin. The spectrum narrows by the end of the tank because of wave damping. This also hints at a possible interpretation of the dips as the transient appearance of a gray soliton after the depth change: the quantitative clarification of their nature will be addressed in a future work. 

Extending the numerical simulations to a propagation distance of $x=500$~m reveals these phenomena more clearly than is possible within the experimentally accessible domain. 
\begin{figure}[ht]
\centering
\includegraphics[width=1.0\linewidth]{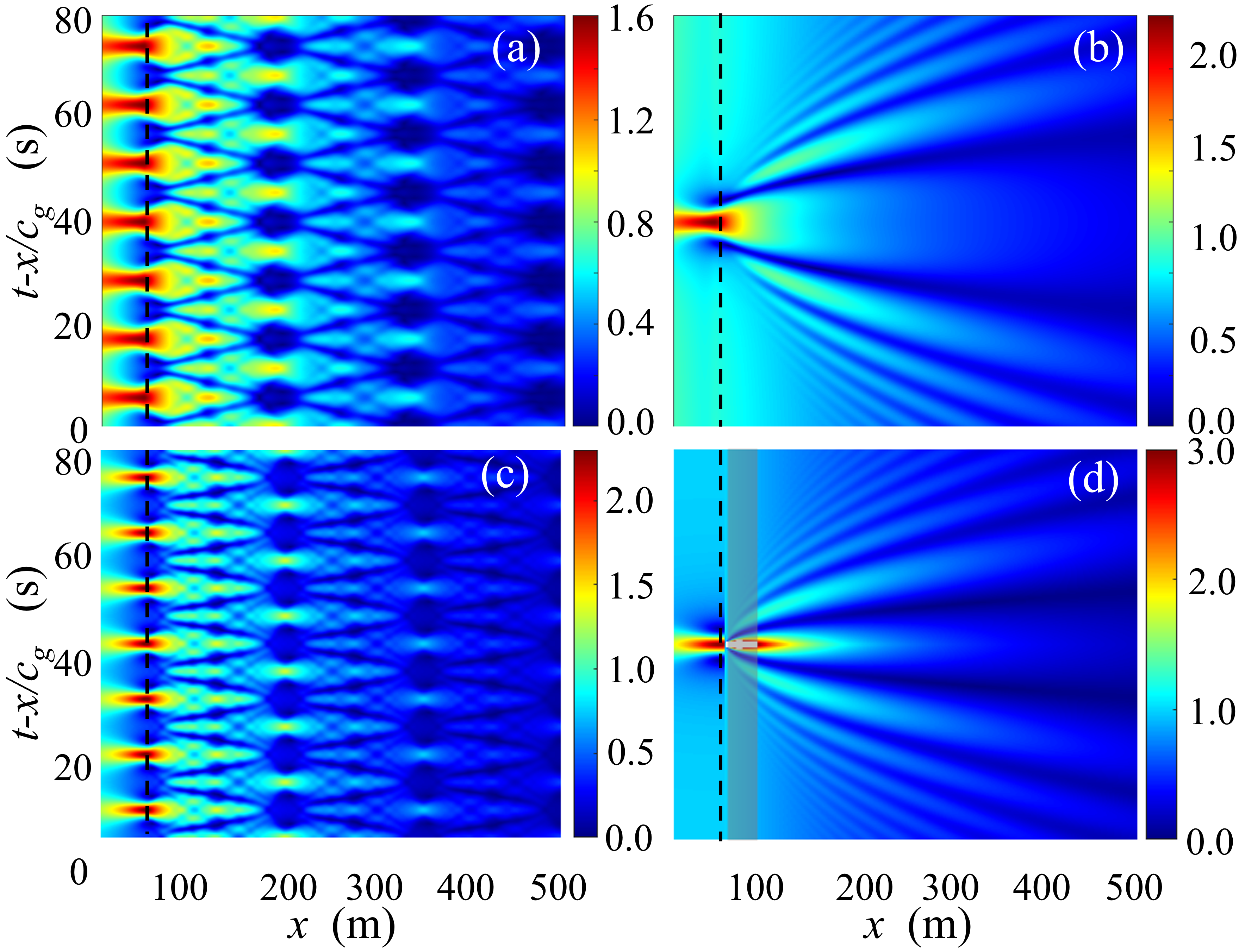}
\caption{Comparison of FDNLSE simulation (a)-(b) showing the long-range evolution of an AB and a PB breather with the linearized theory presented in \hyperref[sec:methods]{Methods}, (c)-(d), normalized by the respective carrier amplitude, over three times the propagation distance in the wave flume. (a) and (c) The same AB case for $\mathfrak{a}=0.25$ as in Fig. \ref{fig2} (b) and (d) the same PB case for the carrier steepness $\varepsilon=0.09$  as in Fig. \ref{fig3}. In (c)-(d) we assume that, after the transition, only the linear dispersion $\beta_2$ is considered, and the approximate solution of the \hyperref[sec:methods]{Methods} section is used. The bathymetry is assumed to jump at $x=50$ m, while the nonlinear solutions are prepared to reach their apex at $x=55$ m. (c) AB, reconstructed with 9 Fourier components, which provides sufficient accuracy.
 (d) PB, based stationary phase approximation; notice that this is valid only for large $x$ and diverges near the transition. The vertical lines share the same meaning as in Fig.~\ref{fig2}.}
\label{fig5}
\end{figure} 
In Fig.~\ref{fig5}(a), we confirm that the AB breaks into an interfering fringe pattern, where the initial expansion is followed by a constructive interference forming a new periodic temporal train of pulses similar to the AB maximal focusing waveform, i.e., a form of recurrence or self-image of the conditions at the depth transition. The process repeats itself, and at each recurrence the maxima are shifted by a half period. This closely resembles a\ac{n initialization of a} Talbot carpet, which, however, becomes weaker and weaker owing to wave attenuation. \ac{Conservative FDNLSE simulations, i.e., with $\mathfrak D=0$, over longer propagation distances would reveal the complete Talbot carpet, characterized by its periodic recurrences.} This setup and transition phenomenon add to the possibilities for Talbot carpet generation in nonlinear dispersive media, as previously reported, for instance, in \cite{bakman2019observation,schiek2021excitation,rozenman2022periodic}. Notice that the simplest Talbot carpet is generated by a diffraction grating, i.e., a mask with periodically spaced slits, corresponding to a square-wave transmission. It was shown that the fringe pattern is fractal between successive recurrences \cite{Berry1996Talbot}. The present situation is more involved, not only because of wave damping, but most importantly because the wave packet at the beginning of the Talbot pattern differs from a pure grating, both in amplitude and phase. The AB exhibits a constant background, and except at the maximal focusing position, where the envelope can be considered purely real, the background is not in phase with the oscillating part. 
As for the PB, we now clearly observe in Fig.~\ref{fig5}(b), that the oscillating tails do not exhibit the typical DSW behavior with a definite speed and the main dips close to the center are damped. Thus, we can interpret the dynamics after the bathymetry change as a pure diffraction-like phenomenon. As with the AB, the background and relative phase of the PB maximum also influence the dynamics, albeit less apparent than in the Talbot pattern, where the diffracting wave components recombine and relative phases matter.

Qualitatively, a rapid depth transition can be considered a step. It is thus worth comparing the simulations of Fig.~\ref{fig5}(a)-(b) with a simplified theory, where the known conditions given by the AB or PB evolution over the initial uniform bathymetry are considered as the initial conditions for the evolution over the final bathymetry region, where $\gamma_2$ is now neglected. This is shown in Fig.~\ref{fig5}(c)-(d). It is clear that the evolution is qualitatively the same for both examples. The main discrepancy is observed in the self-image distance, which turns out to be slightly longer than in the FDNLSE simulations. A linearized theoretical description is therefore sufficient to capture these dynamics, as detailed in \hyperref[sec:methods]{Methods}. This confirms the soundness of the interpretation of the rapid bathymetry change as inducing a time-analogue of diffraction in hydrodynamics.

\section*{Discussion}
We have reported an experimental study investigating the dynamics of breathers transitioning from the focusing to the defocusing regime, which can be achieved in water waves through a rapid reduction in water depth. Our results confirm that the nonlinear focusing process is not compromised during the transition over the steep slope. However, analogous to diffraction from multiple or single slits, the AB and PB exhibit distinct evolution patterns in the defocusing regime. While attenuating in amplitude in the shallower section of the tank, the AB develops a Talbot carpet-like pattern, whereas the PB evolves into two unsteady amplitude dips with oscillatory tails on either side. This latter temporal analogue of diffraction is well described by a linearized theory. Nevertheless, its full development requires long propagation distances, making experimental investigations in water wave facilities with varying bathymetry challenging. Further work is required to isolate the effects of the depth transition on extreme focusing patterns and to investigate the possibility of observing DSW dynamics. The excellent agreement between the observed complex wave dynamics and FDNLSE predictions paves the way for exploring similar transitions in nonlinear wave dynamics across other dispersive media\ac{, including broadband wave fields}.

\section*{Methods} 
\label{sec:methods}

\subsection*{Weakly Nonlinear Wave Theory in Variable Water Depth}
The fundamental framework employed is the time-like finite water depth NLSE (FDNLSE) governing the evolution of wave envelope $\psi$, as derived from the Euler equations for free surface flows via the method of multiple scales \cite{hasimoto1972nonlinear,mei2005theory}
\begin{equation}
i\left(\pder{\psi}{x}+\beta_1\pder{\psi}{t}\right)=
\frac{1}{2}\beta_2\pdern{\psi}{ t}{2}+\gamma\left|\psi\right|^2\psi -i{\mathfrak{D}}\psi, 
\label{nlse}
\end{equation}
where, from the dispersion relation $\omega=\sqrt{gk\sigma}$, with $\sigma\equiv \tanh{\kappa}$, $\kappa\equiv k h$, $k$ the wavenumber, $h$ the water depth and $g$ the gravitational acceleration, we derive
\begin{align*}
\beta_1\equiv &\frac{1}{c_g}=\left\{\frac{g}{2\omega}\left[\sigma + \kappa(1-\sigma^2)\right]\right\}^{-1}\\
\beta_2\equiv &\dfrac{\partial^2 k}{\partial \omega^2}=-\dfrac{1}{c_g^3}\dfrac{\partial^2\omega}{\partial k^2} =\frac{1}{\omega c_g}\left[1-\frac{gh}{c^2_g}(1-\kappa\sigma)(1-\sigma^2)\right],
\end{align*}
Moreover, the nonlinear coefficient reads
\begin{align*}
\gamma
& 
\equiv \frac{\omega k^2}{16\sigma^4 c_g}\left\{9-10\sigma^2+9\sigma^4\right.
\\
&-\left.\frac{2\sigma^2c^2_g}{gh-c_g^2}\left[4\frac{c^2_p}{c^2_g}+4\frac{c_p}{c_g}(1-\sigma^2)+\frac{gh}{c^2_g}(1-\sigma^2)^2\right]       \right\},
\end{align*}
with $c_g\equiv \pder{\omega}{k}$ and $c_p \equiv \frac{\omega}{k}$ denoting group and phase velocity of the carrier wave, respectively. Note that, for simplicity, both the experimental data and the corresponding theoretical predictions are presented in a reference frame moving at the local group velocity $c_g$. Moreover, although more sophisticated wave evolution models, such as that proposed in \cite{DjordjevicRedekopp1978}, could have been employed, we retain the classical NLSE because it provides a simpler framework for interpreting the results. Furthermore, the shoaling corrections are expected to have only a minor influence on the wave evolution under the present experimental conditions: given the transition from initial depth $h_1$ to final depth $h_2$, the shoaling coefficient is estimated as $K_s=\sqrt{c_\mathrm{g1}/c_\mathrm{g2}}$ and thus amounts here to $K_s \approx 0.92$. That being said, a phenomenological attenuation coefficient $\mathfrak{D}$ is introduced, as determined from the data. The latter depends on friction between the fluid and the tank’s bottom and sidewalls, as well as on the carrier wave period, and becomes increasingly important for long tank experiments \cite{kimmoun2016modulation}. The carrier amplitude $a$ and carrier frequency $f=\frac{\omega}{2\pi}$ are first set in the experiments. Given the depth $h$, all other related coefficients are computed accordingly. Apart from the damping, Eq.~\eqref{nlse} is a standard NLSE, which can describe two distinct wave propagation regimes in uniform water depth, depending on the dimensionless depth parameter $\kappa$. In fact, if $\kappa>1.363$, then $\beta_2\gamma>0$, the NLSE is denoted as focusing, and MI, bright solitons, and breathers are admitted. If $\kappa<1.363$, then $\beta_2\gamma<0$, the FDNLSE is denoted as defocusing and admits families of dark solitons, i.e., black and gray, but MI is suppressed \cite{shabat1972exact,akhmediev1997nonlinear,kharif2008rogue,osborne2010nonlinear}. We emphasize that, contrary to optical Kerr media, only the sign of the nonlinearity coefficient $\gamma$ changes in the transition from the focusing to the defocusing regime (and vice versa) at $\kappa=1.363$. We refer to \cite{chabchoub2016hydrodynamic} for details on dimensional transformations of the solutions of the dimensionless NLSE in order to apply these to the water wave problem, particularly, the AB and PB solutions of interest in this study. A description of the boundary conditions to be generated and applied to the wave maker can also be found in the latter reference.

\subsection*{Linear Schr\"odinger framework}\label{sec:end-matter}

The purpose of this analysis is to demonstrate that the essential features of the observed patterns can be explained by assuming that the dynamics beyond the focusing point are predominantly governed by linear dispersive effects. Working for the time being in \textit{dimensionless} units, we assume that the apex or focus point of both the Peregrine breather (PB) and the Akhmediev breather (AB) occurs at the distance $x=x_0$ and that the bathymetry discontinuously jumps at $x=x_j\neq x_0$. For $x>x_j$ the nonlinearity is assumed to be vanishing, while the dispersion value appears only when the comparison to simulation results in dimensional units. We conjecture that a good approximation is to assume that the propagation beyond this point is dominated by linear dispersive effects. In other words, we assume that the envelope evolves according to the {\em linear} Schr\"odinger equation
\begin{eqnarray} \label{LS}
i \frac{\partial \psi}{\partial x} =  \frac{\bar\beta_{2}}{2} \frac{\partial^2 \psi}{\partial t^2}  
\end{eqnarray}
subject to the initial condition $\psi_0=\psi(t,x_j)$ at $x=x_j$ that may or may not correspond to the apex of the PB or the AB. The variables are dimensionless, i.e., $\psi\to a \psi$, so that the breather background before the bathymetry change $\psi_0 =1$, $t\to T_0 t$ and $x\to X_0 x$, where $X_0= \left[\gamma^{(1)} a^2\right]^{-1} $ and $T_0 \equiv \left[X_0\beta_2^{(1)} \right]^{\frac{1}{2}}$, the superscripts refer to the value of parameters before the bathymetry change and $a$ is the initial carrier amplitude, like above. It finally turns out that $\bar\beta_{2}=\beta_2^{(2)}/\beta_2^{(1)}$.  

Introducing the shifted coordinate $\hat{x}=x-x_j$ and the Fourier transform and inverse transform
\begin{equation}
\begin{aligned}
\tilde \psi(\omega,\hat{x}) &= \int_{-\infty}^{+\infty} \psi(t,\hat{x}) \exp(i \omega t) \text{d}t; \\ 
\psi(t,\hat{x}) &= \frac{1}{2\pi} \int_{-\infty}^{+\infty} \tilde \psi(\omega,\hat{x}) \exp(-i \omega t) \text{d}\omega,
\end{aligned}
\end{equation}
the solution of Eq.~\eqref{LS} can be immediately expressed in Fourier space as
\begin{eqnarray} \label{Fsol}
\tilde \psi(\omega,\hat{x})  = \tilde \psi_0(\omega,\hat{x}=0) \exp \left(i \frac{\bar\beta_2 \omega^2}{2} \hat{x} \right),
\end{eqnarray}
which then yields the solution for $\hat{x} \ge 0$ as 
\begin{eqnarray} \label{integral}
\psi(t,\hat{x})  = \frac{1}{2\pi} \int_{-\infty}^{+\infty} \tilde \psi_0(\omega,0) e^{i \frac{\bar\beta_2 \omega^2}{2} \hat{x}} e^{-i\omega t} \text{d}\omega.
\end{eqnarray}
Equation \eqref{integral} has the characteristic form of a diffraction integral in the Fresnel approximation. 

\subsection*{Linear Evolution of Akhmediev Breather}As far as the AB is concerned, a slightly different approach is required. Neglecting the global phase factor, we have at the jump position $x=x_j$ the following periodic function with temporal period $2\pi/\omega_0$, where $\omega_0=2\sqrt{1-2\mathfrak{a}}$, along with its formal Fourier expansion
\begin{align}
\psi_{AB}(t,\hat{x}=0)
&= 1 + \frac{
2(1-2\mathfrak{a})\cosh(\mathfrak{b}\tilde{x})
+ i\mathfrak{b}\sinh(\mathfrak{b}\tilde{x})
}{
\sqrt{2\mathfrak{a}}\cos(\omega_0 t)
- \cosh(\mathfrak{b}\tilde{x})
}
\nonumber\\
&= \sum_{n=-\infty}^{+\infty} \tilde{u}_{n} e^{i n \omega_0 t}.
\end{align}
with $\tilde x\equiv x_j-x_0$ and $\mathfrak{b}=\sqrt{8\mathfrak{a}(1-2\mathfrak{a})}$. 
The Fourier coefficients $\tilde{u}_{n}$ are obtained by conventional techniques and read
\begin{align} \label{pumpAB}
\tilde{u}_{0}&=1-\frac{C(\tilde x)}{g(\tilde x)};\\
\tilde{u}_n &=-\frac{C(\tilde x)}{g(\tilde x)}\rho(\tilde x)^{|n|},
\end{align}
with $C(x) \equiv 2(1-2\mathfrak{a})\cosh{\mathfrak{b} x} + i\mathfrak{b} \sinh{\mathfrak{b} x}$, $g(x) = \sqrt{\cosh^2{\mathfrak{b} x}-2\mathfrak{a}}$ and $\rho(x)\equiv \frac{\cosh{\mathfrak{b} x}-g}{\sqrt{2\mathfrak{a}}}$.
The term $n=0$ corresponds to the carrier wave, and $n \neq 0$ for the symmetric ($\pm n$) sideband pairs. They reduce to those reported in Ref.~\cite{Bendahmane2015} for $\tilde x = 0$.
In this case, the evolution of the field is such that each discrete frequency pair $\pm n \omega_0$ undergoes a phase rotation given by the term $\exp \left[i \frac{\bar\beta_2}{2} (n \omega_0)^2 \hat{x} \right]$ induced by dispersion. Therefore, we obtain the evolving field from the following formula, where we exploited the parity of the coefficients of order $\pm n$ 
\begin{eqnarray} \label{ABevolution}
\psi(t,\hat{x})  =  \tilde u_0 + 2 \sum_{n=1}^{\infty} \tilde u_n \cos(n \omega_0 t)
~e^{ i \frac{\bar\beta_2 n^2 \omega_0^2}{2} \hat{x}}
\end{eqnarray}
The result is that, in the linear limit, the AB gives a carpet-like evolution as shown in Fig.~\ref{fig5}(a,c), where the initial pattern is recovered at distances $\hat{x}_m = m 2\pi/(\beta \omega_0^2)$, with even $m=0,2,\ldots$, whereas $\pi$-shifted patterns are generated for odd $m=1,3,\ldots$.

\subsection*{Linear Evolution of Peregrine Breather}We study the PB first, because it is a direct application of Eq.~\eqref{integral}.
Its envelope at the jump position reads
\begin{equation}    
\label{eq:PB}
\psi_{PB}(t,\hat{x}=0)  = 1-4\frac{1- 2 i \tilde{x}}{1+4t^2 + 4\tilde{x}^2},
\end{equation} 
with $\tilde x\equiv x_j-x_0$. The 
 Fourier transform associated to Eq.~\eqref{eq:PB} reads
\begin{equation} \label{PBfourier}
 \tilde \psi_0=\tilde \psi(\omega,0)  = 2\pi \delta (\omega) - 2\pi e^{i\phi_A}\exp(-|A\omega|/2),
 \end{equation}
with $A \equiv 1-2i \tilde x=|A|e^{i\phi_A}$.

We substitute $\tilde\psi(\omega,0)$ from Eq.~\eqref{PBfourier} into Eq.~\eqref{integral} to obtain
\begin{eqnarray} \label{PSint}
\psi(t,\hat{x})  = 1- e^{i\phi_A}\int_{-\infty}^{+\infty} \chi(\omega) e^{i \phi(\omega)} \text{d}\omega
\end{eqnarray}
where $\chi(\omega)=e^{-\frac{|A\omega|}{2}}$ is a rapidly decaying real function, and $\phi(\omega)=\left(\frac{\bar\beta_2 \omega^2}{2} \hat{x}-\omega t \right)$ is the phase that rules the fast oscillating part of the integral. In principle, the integral in Eq.~\eqref{PSint} could be expressed in terms of the Faddeeva function. A more transparent result, in the same spirit used for the semiclassical approximation of the Schr\"odinger equation, is obtained by applying the stationary phase approximation \cite{Miller2006asymbook}. In this framework, the dominant contribution is given by the point where the phase is stationary $\phi'(\omega)=0$, i.e., the frequency $\omega =\frac{t}{\bar\beta_2 \hat x}$, and the final result reads 

\begin{align}
\psi(t,\hat x)  =& 1 - \sqrt{\frac{2\pi}{\bar\beta_2 \hat x}} \exp \left(i \frac{\pi}{4} {\rm sgn}(\bar\beta_2 \hat x) + i\phi_A \right)\nonumber     \\
&\times\exp \left(- \left| \frac{A t}{2\bar\beta_2 \hat x} \right| \right) 
\exp \left(- i \frac{t^2}{2 \bar\beta_2 \hat x}  \right),\; \text{for} \;\hat x\to \infty
\label{statphase}
\end{align}

The pattern predicted by Eq.~\eqref{statphase}, shown in Figs. \ref{fig3} and \ref{fig5}, closely resembles the observed pattern when starting from a PB.


\bibliography{Refs}

\section*{Acknowledgements}
A.A.~acknowledges the financial support of the University of Ferrara (Grant FIRD 2024). 
A.A.~and S.T.~acknowledge the financial support of Italian MUR (Ministero dell’Università e della Ricerca) 
under grants 2020X4T57A and 2022NCTCY -NextGenerationEU, with the European Union. J.Z.~acknowledges the financial support of Fundamental Research Funds for the Central Universities (Grant No. 79000039). H.C.H.~acknowledges the financial support of National Science and Engineering Council of Taiwan, Republic of China (MOST 110-2221-E-110-016-MY3). L.K. and M.B.~acknowledge support from the French-Taiwanese ORCHID Program of the Hubert Curien Partnership Program (PHC). A.C.~acknowledges support from Okinawa Institute of Science and Technology (OIST) with subsidy funding from the Cabinet Office, Government of Japan.

\marginpar{\footnotesize }

\section*{Data availability}
All data supporting the findings of this study are available from the corresponding
author upon reasonable request.
\section*{Code availability}
The codes for the analysis of this study are available from the corresponding author
upon reasonable request.

\section*{Author contributions}
A.C. conceived the study. H.-C.H, M.-S.L., L.K, J.Z., and A.C. performed the experiments. A.A., J.Z., and A.C. carried out the numerical simulations. A.A. and S.T. developed the Fresnel approximation for the breather evolution. All authors analyzed the data, interpreted the results, and contributed to the writing and revision of the manuscript.
\end{document}